\documentclass[11pt]{article}
\usepackage{amssymb}
\usepackage{amsmath}

\catcode`@=11
\renewcommand{\section}{\@startsection{section}{1}{0pt}{\medskipamount}
{\medskipamount}{\large\bf}}
\catcode`@=12

\def\a{\alpha}

\def\th{\theta}

\newcommand{\C}{\mathbb C}
\newcommand{\R}{\mathbb R}
\newcommand{\Z}{\mathbb Z}
\newcommand{\N}{\mathbb N}
\newcommand{\Hcal}{{\cal H}}

\def\e{\mbox{e}}
\def\i{\mbox{i}}
\def\N2{$N{=}2$}
\def\pa{\mbox{$\partial$}}
\def\diff{\mbox{d}}
\def\tr{{\rm tr}}
\def\sfrac#1#2{{\textstyle\frac{#1}{#2}}}
\def\rd#1{\buildrel{_{_{\hskip 0.01in}\rightarrow}}\over{#1}}
\def\ld#1{\buildrel{_{_{\hskip 0.01in}\leftarrow}}\over{#1}}

\newcommand{\adag}{a^{\dagger}}

\newcommand{\fh}{\hat{f}}
\newcommand{\gh}{\hat{g}}

\newcommand{\zb}{\overline{z}}

\def\>{\rangle}
\def\<{\langle}

\begin{document}
\begin{titlepage}
\setcounter{page}{0}
\begin{flushright}
hep-th/0108118\\
ITP--UH--23/01\\
\end{flushright}

\vskip 2.0cm

\begin{center}

{\Large\bf  

Scattering of Noncommutative Solitons in 2+1 Dimensions~${}^+$

}

\vspace{14mm}

{\large Olaf Lechtenfeld\ and\ Alexander D. Popov~$^*$ }
\\[5mm]
{\em Institut f\"ur Theoretische Physik  \\
Universit\"at Hannover \\
Appelstra\ss{}e 2, 30167 Hannover, Germany }\\
{Email: lechtenf, popov@itp.uni-hannover.de}

\end{center}

\vspace{2cm}

\begin{abstract}

\noindent
Interactions of noncommutative solitons in a modified $U(n)$ sigma model 
in $2{+}1$ dimensions can be analyzed exactly.  Using an extension of 
the dressing method, we construct explicit time-dependent solutions of 
its noncommutative field equation by iteratively solving linear equations. 
The approach is illustrated by presenting bound states and right-angle 
scattering configurations for two noncommutative solitons.

\end{abstract}

\vfill

\textwidth 6.5truein
\hrule width 5.cm
\vskip.1in

{\small
\noindent ${}^*$
On leave from Bogoliubov Laboratory of Theoretical Physics, JINR,
Dubna, Russia\\
${}^+$
Work partially supported by DFG under grant Le 838/7-1}

\end{titlepage}

\section{Introduction}

\noindent
Noncommutativity of spatial coordinates offers a way to introduce nonlocality
into (quantum) field theory without losing control over its structure.
Before attempting to quantize noncommutative field theories, it is certainly
warranted to characterize the moduli space of their classical configurations.
The generalization of solitons and instantons
to the noncommutative realm has just begun. 
For recent reviews on the subject and its connection with string 
and brane theory see~\cite{nekrasov,komaba,douglas,konechny} 
which include the original literature.

In ~\cite{LPS1,LPS2} we have shown that open \N2 strings in a 
$B$-field background induce on the world volume of $n$ coincident D2-branes a
noncommutative generalization of a modified $U(n)$ sigma model. 
The topological nature of \N2 strings and the integrability of their tree-level
dynamics render this noncommutative sigma model integrable. 
In~\cite{LPS2,LPj} we have described a family of multi-soliton solutions to 
its noncommutative field equation.  These multi-soliton configurations, 
explicitly given for finite values of the noncommutativity parameter~$\th$  
and for any $U(n)$ gauge group, represent $q$~lumps of energy 
moving with constant velocities and escaping completely unharmed from 
their mutual encounters, i.e. they do not interact. 

The question arises whether the no-scattering feature of these solitons is
a consequence of the integrability of the theory or merely due to the 
restrictions we imposed on the form of the solutions. 
This paper affirms the second explanation: 
We show how to explicitly construct noncommutative multi-soliton configurations 
with nontrivial interaction within the modified sigma model, 
generalizing the class of solutions presented in~\cite{LPS2,LPj}.

The no-scattering configurations were obtained 
with a (noncommutative variant of) a solution-generating technique 
called the `dressing method'~\cite{zakharov,zakh2,forgacs,Uh,BB},
by using for the $\psi$ function an ansatz with only first-order poles in
the spectral parameter~$\zeta$. 
Here, our goal is to extend this approach to the case where
$\psi$ possesses higher-order poles.
We shall see that this property leads to genuine soliton-soliton interaction.

After presenting the model and our conventions concerning noncommutativity, we 
briefly review the dressing approach, specialize it to the cases of interest,
and reduce it to a couple of linear equations. 
As examples we construct two kinds of (abelian and nonabelian) solutions: 
(a) a two-soliton bound state exhibiting a time-dependent ring-like shape and
(b) two solitons colliding head-on and scattering at $90^\circ$.
To our knowledge these are the first {\it exact\/} finite-$\th$ multi-soliton
scattering configurations in $2{+}1$ dimensions.

\section{Noncommutative modified sigma model in 2+1 dimensions}

\noindent
{\bf Commutative model.}
As has been known for some time, nonlinear sigma models in $2{+}1$ dimensions
may be Lorentz-invariant or integrable but not both.
In this paper we choose the second property and investigate the
noncommutative extension of a modified $U(n)$ sigma model
(so as to be integrable) introduced by Ward~\cite{ward}.
This model describes the dynamics of a $U(n)$-valued field~$\Phi(t,x,y)$
whose arguments are the (real) coordinates $(t,x,y)$ of~$\R^{1,2}$
with Minkowski metric~$\textrm{diag}(-1,+1,+1)$.

The commutative field equation for this modified sigma model reads
\begin{equation} \label{comuteq}
-\pa_t\,(\Phi^{-1}\pa_t\Phi)+\pa_x\,(\Phi^{-1}\pa_x\Phi)
+\pa_y\,(\Phi^{-1}\pa_y\Phi)
+\pa_y\,(\Phi^{-1}\pa_t\Phi)-\pa_t\,(\Phi^{-1}\pa_y\Phi)\ =\ 0 \quad,
\end{equation}
the last two terms of
which explicitly break the Lorentz group of $SO(1,2)$ to the $GL(1,\R)$
generated by the boost in $y$~direction. Nevertheless, the model possesses
a conserved and rotationally invariant energy functional,
\begin{equation} \label{comutenergy}
E\ =\ \int\!\diff{x}\,\diff{y}\;{\cal E}\ =\ 
\frac{1}{2}\int\!\diff{x}\,\diff{y}\;\tr\,
\Bigl(\pa_t \Phi^{\dagger} \pa_t \Phi \,+\,
 \pa_x \Phi^{\dagger} \pa_x \Phi \,+\,
 \pa_y \Phi^{\dagger} \pa_y \Phi \Bigr) \quad,
\end{equation}
where `tr' implies the trace over the $U(n)$ group space, 
and $\Phi^\dagger=\Phi^{-1}$.

The field equation~(\ref{comuteq}) can be obtained from the self-dual
Yang-Mills equations in~$\R^{2,2}$ by dimensional reduction and a 
(non-covariant) choice of gauge. 
In this interpretation, $\Phi$ arises as a (Yang-type) prepotential~\cite{Y} 
for the Bogomolnyi equations of the $2{+}1$ dimensional
Yang-Mills-Higgs system~\cite{LPj}.

\noindent
{\bf Noncommutative model.}
The noncommutative extension of this classical field theory is achieved by
deforming the ordinary product of classical fields (or their components)
to the noncommutative star product
\begin{equation}
(f \star g)(t,x,y)\ =\ f(t,x,y)\,\exp\,\bigl\{ \frac{\i}{2}
( {\ld{\partial}}_x \,\theta\, {\rd{\partial}}_y
- {\ld{\partial}}_y \,\theta\, {\rd{\partial}}_x \bigr\}\,g(t,x,y)\quad,
\end{equation}
where $\th$ is a positive real constant.
Note that the time coordinate remains commutative.
In the convenient coordinate combinations
\begin{equation} \label{lightcone}
u\ :=\ \sfrac{1}{2}(t+y)\quad,\qquad
v\ :=\ \sfrac{1}{2}(t-y)\quad,\qquad
\pa_u\ =\ \pa_t+\pa_y\quad,\qquad
\pa_v\ =\ \pa_t-\pa_y
\end{equation}
the noncommutative field equation becomes
\begin{equation} \label{yangtype}
\pa_x\,(\Phi^{-1}\star\pa_x\Phi)-\pa_v\,(\Phi^{-1}\star\pa_u\Phi)\ =\ 0 \quad.
\end{equation}

The nonlocality of the star product renders explicit computations cumbersome.
We therefore pass to the operator formalism,
which trades the star product for operator-valued spatial coordinates
$(\hat{x},\hat{y})$ or their complex combinations
$(\hat{z},\hat{\zb})=(\hat{x}+\i\hat{y},\hat{x}-\i\hat{y})$,
subject to
\begin{equation}
[t,\hat{x}]\ =\ [t,\hat{y}]\ =\ 0\quad, \qquad
[\hat{x},\hat{y}]\ =\ \i\theta
\qquad\Longrightarrow\qquad
[\hat{z},\hat{\zb}]\ =\ 2\theta \quad.
\end{equation}
The latter equation suggests the introduction of
creation and annihilation operators,
\begin{equation} \label{adef}
a\ =\ \frac{1}{\sqrt{2\theta}}\,\hat{z} \qquad\textrm{and}\qquad
\adag\ =\ \frac{1}{\sqrt{2\theta}}\,\hat{\zb} \qquad \textrm{with}\quad
[a,\adag]\ =\ 1 \quad,
\end{equation}
which act on a harmonic-oscillator Fock space $\Hcal$ with an orthonormal basis
$\{|n\rangle,\,n=0,1,2,\ldots\}$ such that
\begin{equation}
\adag a\,|n\rangle\ =:\ N\,|n\rangle\ =\ n\,|n\rangle \quad,\qquad
a\,|n\rangle\ =\ \sqrt{n}\,|n{-}1\rangle \quad, \qquad
\adag|n\rangle\ =\ \sqrt{n{+}1}\,|n{+}1\rangle \quad .
\end{equation}

\noindent
{\bf Moyal-Weyl map.}
Any function $f(t,z,\zb)$ can be related to an operator-valued
function $\fh(t)\equiv F(t,a,\adag)$ acting in $\Hcal$,
with the help of the Moyal-Weyl map 
\begin{equation}
f(t,z,\zb)\quad \longrightarrow \quad F(t,a,\adag)\ =\
\textrm{Weyl-ordered} \  f(t,\sqrt{2\th}\,a,\sqrt{2\th}\,\adag) \quad.
\end{equation}
The inverse transformation recovers the c-number function,
\begin{equation} \label{inverseMoyal}
F(t,a,\adag)\ \longrightarrow\ f(t,z,\zb)\ =\ F_\star \Bigl(
t,\frac{z}{\scriptstyle\sqrt{2\th}},\frac{\zb}{\scriptstyle\sqrt{2\th}}
\Bigr) \quad,
\end{equation}
where $F_\star$ is obtained from $F$ by replacing ordinary with star products.
Under the Moyal-Weyl map, we have
\begin{equation}
\pa_z f \quad \longrightarrow \quad \hat{\pa}_z \fh
\ =\ \frac{-1}{\sqrt{2\th}}\,[\adag,\fh]
\qquad\textrm{and}\qquad
\pa_{\zb} f \quad \longrightarrow \quad \hat{\pa}_{\zb} \fh
\ =\ \frac{1}{\sqrt{2\th}}\,[a,\fh] \quad,
\end{equation}
\begin{equation} \label{trace}
f\star g\ \longrightarrow\ \fh\,\gh \qquad\textrm{and}\qquad
\int\! \diff{x}\,\diff{y}\,f\ =\
2\pi \theta \,\mbox{Tr}\, \fh\ =\
2\pi \theta \sum_{n \geq 0} \langle n|\fh |n \rangle \quad,
\end{equation}
where `Tr' signifies the trace over the Fock space~$\Hcal$.
For notational simplicity we will from now on omit the hats over the operators
except when confusion may arise.

\section{Dressing approach and explicit solutions}

\noindent
The payoff for considering an integrable model is the availability of
powerful techniques for constructing solutions to the equation of motion.
One of these tools is the so-called `dressing method', which was invented to
generate solutions for commutative integrable systems
\cite{zakharov,zakh2,forgacs,Uh,BB}
and is easily extended to the noncommutative setup~\cite{LPS2,LPj}.
Let us briefly present this method (already in the noncommutative context)
before applying it to the modified sigma model.

\noindent
{\bf Linear system.}
The key observation is that our field equation can be obtained as a
compatibility condition for a linear system.
We consider the two linear equations
\begin{equation}\label{linsys}
(\zeta \pa_x -\pa_u)\psi\ =\ A\,\psi \qquad\textrm{and}\qquad
(\zeta \pa_v -\pa_x)\psi\ =\ B\,\psi \quad,
\end{equation}
where $\psi$ depends on $(t,x,y,\zeta)$ or, equivalently, on $(x,u,v,\zeta)$
and is an $n{\times}n$ matrix whose elements act as operators in the Fock
space $\Hcal$. The matrices $A$ and~$B$ are of the same type as $\psi$ but
do not depend on~$\zeta$.
The spectral parameter~$\zeta$ lies in the extended complex plane.
The matrix $\psi$ is subject to the following reality condition~\cite{ward}:
\begin{equation}\label{real}
\psi(t,x,y,\zeta)\;[\psi(t,x,y,\bar{\zeta})]^{\dagger}\ =\ 1 \quad,
\end{equation}
where `$\dagger$' is hermitean conjugation.
The compatibility conditions for the linear system of differential equations
(\ref{linsys}) read
\begin{align}
\pa_x B -\pa_v A\ =\ 0 \quad ,\label{comp1} \\[4pt]
\pa_x A -\pa_u B -[A,B]\ =\ 0 \quad . \label{comp2}
\end{align}
We can solve the second equation by putting
\begin{equation} \label{Lax2}
A\ =\ \Phi^{-1}\,\pa_u\Phi
\qquad\textrm{and}\qquad
B\ =\ \Phi^{-1}\,\pa_x\Phi \quad,
\end{equation}
which transforms the first equation into (the operator
version of) our Yang-type equation~(\ref{yangtype}),
\begin{equation} \label{yangtype2}
\pa_x\,(\Phi^{-1}\,\pa_x\Phi)-\pa_v\,(\Phi^{-1}\,\pa_u\Phi)\ =\ 0 \quad.
\end{equation}
Inserting the parametrization of $A$ and $B$ into the linear system
(\ref{linsys}) we immediately obtain the standard conditions~\cite{ivle}
\begin{align}
\psi(t,x,y,\zeta\to\infty)\ &=\ 1\ +\ O(\zeta^{-1})
\quad, \label{asymp1} \\[4pt]
\psi(t,x,y,\zeta\to0)\ &=\ \Phi^{-1}(t,x,y)\ +\ O(\zeta)\quad. \label{asymp2}
\end{align}
Note that the second equation yields $\Phi$ directly in terms of~$\psi$
and thus also $A$ and $B$ via~(\ref{Lax2}) or directly from
\begin{align}
-\psi(t,x,y,\zeta)\;(\zeta\pa_x-\pa_u)[\psi(t,x,y,\bar{\zeta})]^{\dagger}\
=\ A(t,x,y) \quad , \label{A1} \\[4pt]
-\psi(t,x,y,\zeta)\;(\zeta\pa_v-\pa_x)[\psi(t,x,y,\bar{\zeta})]^{\dagger}\
=\ B(t,x,y) \quad \label{B1}.
\end{align}

\noindent
{\bf Ansatz.}
Having identified auxiliary linear first-order differential equations
pertaining to our second-order nonlinear equation, we set out to solve
the former.
In two previous papers~\cite{LPS2,LPj} we considered 
solutions $\psi$ of the linear system~(\ref{linsys}) 
containing only first-order poles in $\zeta$,
\begin{equation}
\psi\ =\ 1\ +\ \sum\limits^m_{k=1}\frac{R_k}{\zeta - \mu_k} \quad ,
\end{equation}
where the $\mu_k$ are complex constants and
the $n{\times}n$ matrices $R_{k}(t,x,y)$ are independent of~$\zeta$.
To such $\psi$ there correspond
solutions $A$ and $B$ of equations (\ref{comp1}) and (\ref{comp2}) which are
parametrized by some $n\times r$ matrices $T_k(t,x,y)$~\cite{LPS2,LPj}.

The simplest case occurs when $\psi$ has only one pole at $\zeta =-\i$,
\begin{equation} \label{static}
\psi\ =\ 1\,+\,\frac{R}{\zeta+\i}\ =:\ 1\,-\,\frac{2\i}{\zeta+\i}\,P
\qquad\textrm{so that}\qquad \Phi\ =\ 1\,-\,2\,P \quad.
\end{equation}
In this case all configurations are static and parametrized by a hermitean
projector $P=T\frac{1}{T^\dagger T}T^\dagger$ which obeys
\begin{equation} \label{eomstatic}
(1-P)\,\pa_{\bar z}\,P\ =\ 0 \qquad\Longrightarrow\qquad
\ (1-P)\,a\,P\ =\ 0 \quad .
\end{equation}
Then $T$ satisfies the equation
\begin{equation}
\ (1-P)\,a\,T\ =\ 0 \quad ,
\end{equation}
which means that $aT$ lies in the kernel of $1{-}P$.
Recall that $T$ is an $n\times r$ matrix. In the abelian case $n=1$, and $r$ 
is the rank of the projector $P$ in the Fock space $\Hcal$. 
In the nonabelian case $n\ge 2$, and $r(<n)$ can be identified with 
the rank of the projector in the $U(n)$ group space (but not in $\Hcal$). 
In terms of $P$ the matrices $A$ and $B$ from the linear system~(\ref{linsys}) 
are expressed as
\begin{equation}
A\ =\ -2\i\,\pa_x P \qquad\textrm{and}\qquad
B\ =\ 2\i\,\pa_y P\quad .
\end{equation}

\noindent
{\bf Dressing approach.}
The dressing method is a recursive procedure for generating a new solution
$(\tilde\psi,\tilde A,\tilde B)$ from an old one, $(\psi,A,B)$ 
by multiplication,
\begin{equation} \label{generalansatz}
\tilde\psi(t,x,y,\zeta)\ =\ \chi(t,x,y,\zeta)\,\psi(t,x,y,\zeta)
\qquad\textrm{with}\qquad
\chi\ =\ 1\ +\ \sum_{k=1}^m\frac{R_{k}}{\zeta-\mu_k}\quad.
\end{equation}
We now choose a static configuration $\psi$
as a seed solution and consider a dressing transformation
with $\chi$ being of the same form as~$\psi$,
\begin{equation} \label{psi}
\psi\ \to\ \tilde\psi\ =\ \chi\psi\ =\
(1\,-\,\frac{2\i}{\zeta+\i}\,\tilde P)(1\,-\,\frac{2\i}{\zeta+\i} P)\ =\ 
1\,-\,\frac{2\i}{\zeta+\i} (P+\tilde P)\,-\,\frac{4}{(\zeta+\i )^2} \tilde PP
\ , 
\end{equation}
where $\tilde P$ is some matrix. 
This leads to a configuration of the form
\begin{equation} \label{phi2}
\Phi\ =\ \tilde\psi^{-1}(\zeta{=}0)\ =\ (1-2P)\,(1-2\tilde P) \quad.
\end{equation}
Since both $\psi$ and $\chi$ possess a first-order pole at $\zeta=-\i$
our ansatz for $\tilde\psi$ contains a second-order pole. 
Substituting (\ref{psi}) into the reality condition~(\ref{real}) 
directly yields
\begin{equation}
\tilde P^\dagger\ =\ \tilde P \qquad\textrm{and}\qquad 
\tilde P^2\ =\ \tilde P \quad,
\end{equation}
qualifying $\tilde P$ as a hermitean projector, i.e.
\begin{equation}
\tilde P\ =\tilde T\,\frac{1}{\tilde T^\dagger \tilde T}\,\tilde T^\dagger\ 
\end{equation}
with some $n\times\tilde r$ matrix $\tilde T$.

\noindent
{\bf Linear equations.}
Demanding that $\tilde\psi$ is again a solution of the linear equations
(\ref{linsys}) with some $\tilde A$ and $\tilde B$, we obtain
\begin{align}
\tilde A(t,x,y)\ &=\ 
-\tilde\psi (t,x,y,\zeta )\ (\zeta\pa_x-\pa_u)\ 
[\tilde\psi (t,x,y,\bar\zeta)]^\dagger
\nonumber\\ 
&=\ (1-\frac{2\i}{\zeta+\i}\,\tilde P)\,A\,(1+\frac{2\i}{\zeta-\i}\,\tilde P)\ 
-\ (1-\frac{2\i}{\zeta+\i}\,\tilde P)\,(\zeta\pa_x -\pa_u)\,
(1+\frac{2\i}{\zeta-\i}\,\tilde P)\quad, \label{Atxy}
\end{align}
\begin{align}
\tilde B(t,x,y)\ &=\ 
-\tilde\psi (t,x,y,\zeta )\ (\zeta\pa_v-\pa_x)\ 
[\tilde\psi (t,x,y,\bar\zeta)]^\dagger
\nonumber\\ 
&=\ (1-\frac{2\i}{\zeta+\i}\,\tilde P)\,B\,(1+\frac{2\i}{\zeta-\i}\,\tilde P)\
-\ (1-\frac{2\i}{\zeta+\i}\,\tilde P)\,(\zeta\pa_v -\pa_x)\,
(1+\frac{2\i}{\zeta-\i}\,\tilde P) \quad .\label{Btxy}
\end{align}
The poles at $\zeta =\pm\i$ on the r.h.s. of these equations have to be
removable since $\tilde A$ and $\tilde B$ are independent of $\zeta$.
Putting to zero the corresponding residues, we obtain the essential equations
\begin{equation}\label{tp}
(1-\tilde P)\,\left\{\pa_{\bar z}\tilde P\ +\ (\pa_{\bar z} P)
\tilde P\right\}\ =\ 0
\qquad \textrm{and}\qquad
(1-\tilde P)\,\left\{\sfrac{\i}{2}\,\pa_{t}\tilde P + 
(\pa_z P)\tilde P\right\}\  =\ 0\quad .
\end{equation}
With the help of the identities
\begin{equation}
(1-\tilde P)\tilde P\ \equiv\ 0\qquad\textrm{and}\qquad 
(1-\tilde P)\tilde T\ \equiv\ 0
\end{equation}
the equations (\ref{tp}) reduce to
\begin{equation}\label{ppp}
(1{-}\tilde P)\left\{a\tilde T + [a,P]\,\tilde T\right\}
\frac{1}{\tilde T^\dagger\tilde T}\, \tilde T^\dagger = 0
\qquad\textrm{and}\qquad
(1{-}\tilde P)\, \left\{\pa_t\tilde T - 
\i\gamma\,[a^\dagger ,P]\,\tilde T\right\}
\frac{1}{\tilde T^\dagger\tilde T}\,\tilde T^\dagger = 0\ ,
\end{equation}
where $\gamma =-\sqrt{\frac{2}{\th}}$. 
Obviously, a sufficient condition for a solution is
\begin{equation}\label{suff}
a\tilde T \  +\ [a,P]\;\tilde T\ =\ \tilde T\,Z_1
\qquad\textrm{and}\qquad
\pa_t\tilde T \ - \ \i\gamma 
[a^\dagger ,P]\;\tilde T\ = \ \tilde T\,Z_2\quad ,
\end{equation}
with some functions $Z_1(t,a,a^\dagger )$ and $Z_2(t,a,a^\dagger )$.
In the nonabelian case we choose $Z_1=a$ and $Z_2=0$ while 
in the abelian case we take $Z_1=z_1$ and $Z_2=z_2$ with $z_{1,2}\in\C$. 
In both these cases the compatibility conditions for 
the linear equations (\ref{suff}) lead to the equation
\begin{equation}\label{sigma}
[a^\dagger, [a\,,P]\,]\ +\ [\,[a\,,P]\,,[a^\dagger ,P]\,]\ =\ 0 \quad,
\end{equation}
which is the field equation of the two-dimensional Euclidean sigma model
\cite{LPj}. Since our $P$ satisfies this equation, the equations (\ref{suff})
are compatible. 
After constructing a projector $\tilde P$ by solving (\ref{suff}),
one obtains a solution $\tilde\psi$ of the linear system (\ref{linsys}) 
with $\tilde A$ and $\tilde B$ obeying the field equations 
(\ref{comp1}) and~(\ref{comp2}).
Then one can choose $\tilde\psi$, $\tilde A$, $\tilde B$ as a new seed 
configuration and repeat the dressing transformation $\tilde\psi\to\psi'$,
again obtaining linear equations on some matrix $T'$.
This iterative dressing procedure enables one to construct 
various solutions of the field equations.

\section{Nonabelian multi-soliton configurations}

\noindent
{\bf Equations.} 
In order to generate some examples of nonabelian multi-solitons 
we specialize to $U(2)$ and take as a seed configuration the simplest 
nontrivial solution of (\ref{eomstatic}), viz. 
\begin{equation}
P\ =\ T\,\frac{1}{T^\dagger T}\,T^\dagger\ =\ 
\begin{pmatrix} \frac{1}{1+\bar z z} &  \frac{1}{1+\bar z z} \bar z  \\[4pt]
              z \frac{1}{1+\bar z z} & z\frac{1}{1+\bar z z} \bar z 
\end{pmatrix}
\qquad {\mbox{with}}\qquad T={1\choose {z}}\qquad {\mbox{and}}\qquad
\bar z\equiv z^\dagger\quad .
\end{equation}
We build the dressing factor~$\chi$ with a matrix $\tilde T={u\choose {v}}$,
where $u(t,z,\zb)$ and $v(t,z,\zb)$ are functions to be determined.
Substituting this ansatz for $\tilde T$ into the first of eqs.~(\ref{suff}) 
with $Z_1=a$, we obtain
\begin{align}
[z,u]\ &=\ \Bigl[\frac{1}{1+\bar z z}\;,\, z\Bigr](u+\bar zv) 
-\frac{2\th}{1+\bar z z}v \quad,
\nonumber \\ \label{v}
[z,v]\ &=\ z\Bigl[\frac{1}{1+\bar z z}\;,\, z\Bigr](u+\bar zv) 
- z\frac{2\th}{1+\bar z z}v \quad.
\end{align} 
{}From these equations it follows that
\begin{equation}
[z\;,\; zu-v]\ =\ 0 \qquad\Longrightarrow\qquad v\ =\ zu-f(t,z)\quad,
\end{equation}
where $f(t,z)$ is an arbitrary function of $t$ and $z$ 
(i.e. it does not depend on $\bar z$).
In a similar vein, the second of eqs.~(\ref{suff}) with $Z_2=0$ reduces to 
\begin{equation}\label{u}
\i\th\,\pa_tu\ =\ \Bigl[\bar z\,,\,\frac{1}{1+\bar z z} \Bigr] (u+\bar zv) \quad.
\end{equation}

\noindent
{\bf Explicit solutions.}
It is not difficult to see that (\ref{v}) and~(\ref{u}) are solved by
\begin{equation}\label{uvf}
u\ =\ 1+\frac{1}{1+\bar z z}\bar z f\quad ,\qquad
v\ =\ z-\frac{1}{1+\bar z z +2\th} f\qquad\textrm{with}\qquad
f\ =\ -2\i\;(t+h(z))\quad,
\end{equation}
where $h$ is an arbitrary meromorphic function of $z$
 (i.e. independent on~$t$). 
Hence, we have
\begin{equation}\label{T}
\tilde T\ \equiv\ {u\choose {v}}\ =\
{1\choose {z}}\ -\ {\bar z\choose {-1}}\frac{2\i}{1+\bar z z +2\th}\,(t+h(z))
\quad.
\end{equation}
For locations given by $t+h(z)=0$, we see that $\tilde T=T$ and thus
$\Phi=(1-2P)^2=1$ degenerates.
One may obtain more general configurations by replacing $T={1\choose{z}}$
with $T={1\choose{g(z)}}$ containing a meromorphic function $g(z)$
and then solving (\ref{suff}) again.
In commutative limit this family of solutions 
coincides with the ones studied in~\cite{ward2,Io,ioan2}. 
For this reason we will not discuss them in detail.

Depending on the explicit form of $h(z)$, the solutions (\ref{T}) describe
different kinds of multi-soliton configurations. 
Let us take, for example, $h(z)=z^q$ with $q\in\Z$. 
Then, $q\le 1$ leads to a time-dependent ring-like structure 
(for the energy density) while $q\ge 2$ creates a configuration consisting of
$q$ lumps. The latter simultaneously accelerate towards the origin ($z=0$) 
of the noncommutative plane, scatter at an angle of $\pi/q$, and decelerate 
as they separate again (cf.~\cite{ward2,Io,ioan2}).

\noindent
{\bf Examples.} 
To visualize the field configurations we employ the inverse Moyal-Weyl 
transformation~(\ref{inverseMoyal}) and obtain
\begin{equation}
T\ \to\ T_\star \ =\ {1\choose{z}} \qquad\textrm{and}\qquad
\tilde T\ \to\ \tilde T_\star \ =\ {1\choose{z}}\ -\ {\bar z\choose{-1}}\star
\frac{2\i}{1+z\bar z +\th } \star (t+h(z))
\end{equation}
with commutative coordinates $t,z,\bar z$.
Using these expressions we can calculate the projectors 
$P_\star$ and $\tilde P_\star$ as well as the field 
$\Phi_\star=(1-2 P_\star)\star(1-2\tilde P_\star)$ and 
the energy density~${\cal E}_\star$ for this family of solutions. 
Moreover, for large $r^2\equiv z\bar z$ the field $\Phi_\star$ 
and the energy density $\cal E_\star$ will approach their commutative 
limits $\Phi$ and $\cal E$, respectively.
Therefore, the asymptotic analysis of the papers~\cite{ward2,Io,ioan2} 
can be applied without alteration.

The simplest scattering case occurs for $h=z^2$. 
One can show that for large~$r$ the energy density
\begin{equation}
{\cal E}_\star\ = 16\ \frac{1+10r^2+5r^4+4t^2(1+2r^2)-8t(x^2-y^2)}
{(1+2r^2+5r^4+4t^2+8t(x^2-y^2))^2}\ [1+O(\th/r^2)] 
\end{equation}
peaks near the degeneration locus~$z^2+t=0$. 
Hence, for a fixed large (positive or negative) time~$t$, 
two lumps are centered at $z=\pm\sqrt{-t}$. 
Varying~$t$ we see the two lumps accelerating symmetrically towards each other 
along the $x$-axis, interacting at the origin near~$t{=}0$ and decelerating 
to infinity along the $y$-axis. 
Thus, a head-on collision of these solitons results in $90^\circ$ scattering.

To exhibit a configuration with ring-like structure, one can take e.g. $h=0$. 
In this case it turns out that the energy density 
\begin{equation}
{\cal E}_\star\ =\ 16 \frac{1+2r^2+r^4+4t^2(1+2r^2)}
{(1+2r^2+r^4+4t^2)^2}\ [1+O(\th/r^2)] 
\end{equation}
is invariant under rotations
around the origin. For fixed large time~$t$, its degenerate maximum is found at
$r^2\sim 2|t|$. Varying~$t$, this ring shrinks until $t=0$ 
and then expands again.
One can also show that $\Phi_\star(t{\to}\pm\infty)\to-1$, 
i.e. no scattering occurs.

In both these examples, noncommutativity is felt only inside a disc of
radius $R\sim\sqrt{\th}$. For $|t|\gtrsim\th$ then,
the solitons agree qualitatively with their commutative cousins.

\section{Abelian multi-soliton configurations}

\noindent
{\bf Explicit solution.}
The construction of abelian solitons is not achieved by simply taking the
$U(n)$ formalism to $n=1$. The abelian case deserves a special treatment.
For a seed solution, we take the simplest abelian static one-soliton 
configuration, made from the ket $T{=}|0\>$ and described by the projector 
\begin{equation}\label{p}
P\ =\ |0\>\<0|
\end{equation}
satisfying (\ref{eomstatic}). 
The dressing factor~$\chi$ is built from a ket~$\tilde T$ which,
via~(\ref{suff}), is subject to
\begin{equation}\label{tildeT}
a\,\tilde T - |0\>\<1|\,\tilde T\ =\ z_1\,\tilde T\qquad\mbox{and}\qquad
\pa_t\tilde T -\i\gamma\,|1\>\<0|\,\tilde T\ =\ z_2\,\tilde T \quad.
\end{equation}
We put $z_1=z_2=0$ for simplicity.\footnote{
More generally, $z_1$ and $z_2$ characterize the locations of the solitons.}
Substituting a generic 
$\tilde T=|0\>+\sum_{n\ge1}\tilde T_n(t)|n\>$ 
into these equations (the scale drops out), 
we learn that $\tilde T_{n\ge2}=0$ so that one is left with
\begin{equation}\label{T2}
\tilde T\ =\ |0\>\ +\ \tilde T_1(t)\,|1\> \qquad\Longrightarrow\qquad
\tilde T\ =\ |0\>\ +\ \i\gamma t\,|1\>\quad.
\end{equation}
Therefore, we obtain
\begin{equation}\label{tildeP}
\tilde P\ =\ \tilde T\,\frac{1}{\tilde T^\dagger\tilde T}\,\tilde T^\dagger\ =\
\frac{1}{1+\gamma^2 t^2}\Bigl(
|0\>\<0|+\i\gamma t\,|1\>\<0|-\i\gamma t\,|0\>\<1|+\gamma^2t^2|1\>\<1|\Bigr)
\quad.
\end{equation}

{}From the projectors (\ref{p}) and (\ref{tildeP}) we construct 
the field~$\Phi$ \`a la~(\ref{phi2}),
\begin{equation} \label{Phi}
\Phi\ =\ 1\ -\ \frac{2}{1+\gamma^2t^2}
\Bigl( \gamma^2t^2 |0\>\<0| + \i\gamma t\,|0\>\<1|
     + \i\gamma t\,|1\>\<0| + \gamma^2t^2 |1\>\<1| \Bigr) \quad.
\end{equation}
It is easy to see that
\begin{equation}\label{Ph}
\lim_{t\to\pm\infty}\Phi\ =\ 1\ -\ 2\,\Bigl( |0\>\<0|+|1\>\<1| \Bigr)\quad ,
\end{equation}
i.e. both limits coincide with the static two-soliton configuration. 
Therefore, no scattering occurs.

\noindent
{\bf Energy.} 
The energy density for the solution (\ref{Phi}) is computed from
(\ref{comutenergy}),
\begin{align}
{\cal E}\ &=\ 
 \sfrac{1}{2}\pa_t\Phi^\dagger\pa_t\Phi + \pa_z\Phi^\dagger\pa_{\bar z}\Phi
+\pa_{\bar z}\Phi^\dagger\pa_z\Phi 
\nonumber\\[8pt]  &=\
 \sfrac{1}{2}\pa_t\Phi^\dagger\pa_t\Phi  
+\sfrac{1}{2\th} [a^\dagger , \Phi^\dagger ][a^\dagger , \Phi^\dagger ]^\dagger
+\sfrac{1}{2\th} [a , \Phi^\dagger ][a , \Phi^\dagger ]^\dagger
\nonumber\\[8pt]  &=\
 \gamma^2\left\{\frac{2}{(1{+}\gamma^2t^2)^2} \Bigl(|0\>\<0|+|1\>\<1|\Bigr) 
+\frac{4\gamma^2t^2 }{(1{+}\gamma^2t^2)^2}|0\>\<0|
+\frac{2\gamma^2t^2 }{1{+}\gamma^2t^2}\Bigl(|1\>\<1|+|2\>\<2|\Bigr) \right.
\nonumber\\  & \left.\quad
-\frac{\sqrt{2}\,\gamma^2t^2}{(1{+}\gamma^2t^2)^2}\Bigl(|2\>\<0|+|0\>\<2|\Bigr)
+\frac{\i\sqrt{2}\,\gamma^3t^3}{(1{+}\gamma^2t^2)^2}
 \Bigl(\sqrt{2}|1\>\<0|-\sqrt{2} |0\>\<1|+|2\>\<1|-|1\>\<2|\Bigr) \right\}
\quad. \label{calE}
\end{align}
Consequently, the energy of this time-dependent abelian configuration is
\begin{align}
E\ &=\ 2\pi\th\,\mbox{Tr}\,{\cal E}\ =\ 
2\pi\th\gamma^2\left\{ \frac{4}{(1{+}\gamma^2t^2)^2}
+\frac{4\gamma^2t^2}{(1{+}\gamma^2t^2)^2}
+\frac{4\gamma^2t^2}{1{+}\gamma^2t^2} \right\}
\nonumber\\ &=\
 \frac{8\pi\th\gamma^2}{(1{+}\gamma^2t^2)^2}\Bigl\{ 
 1+2\gamma^2t^2+(\gamma^2t^2)^2 \Bigr\}\ =\
 8\pi\th\gamma^2\ =\ 16\pi \quad,\label{E}
\end{align}
where we used the definition $\gamma^2=2/\th$. 
The value of~$16\pi$ is twice the energy of a single static (rank-one) abelian 
soliton, in perfect agreement with the interpretation of~(\ref{Ph}).

\noindent
{\bf Inverse Moyal-Weyl map.}
The translation from operator-valued solutions $\Phi(t)$ 
to functions $\Phi_\star(t,z,\bar z)$ solving the noncommutative field 
equations~(\ref{yangtype}) 
is easily accomplished by the inverse Moyal-Weyl map~(\ref{inverseMoyal}). 
For the configuration~(\ref{Phi}) we obtain
\begin{align}
\Phi_\star\ &=\ 
1\ -\ \frac{8\gamma^4t^2}{1+\gamma^2t^2}\,z\bar z\,\e^{-\gamma^2 z\bar z/2}\
+\ \frac{4\gamma^2t}{1+\gamma^2t^2}\,\i(z+\bar z)\,\e^{-\gamma^2 z\bar z/2}
\nonumber\\  &=\
1\ -\ \frac{8t}{\th +2t^2}\left\{ 2\frac{z\bar z}{\th}\,t - \i(z+\bar z)\right\} 
\,\e^{-z\bar z/\th} \quad, \label{newPhi}
\end{align}
which produces the energy density
\begin{equation}\label{estar}
{\cal E}_\star\ =\ \frac{16\,\e^{-r^2/\th}}{\th^2(1+2t^2/\th )^2}
 \left \{ 2t^2+r^2+\frac{2r^2}{\th}
\Bigl(\frac{r^2}{\th}-1\Bigr)\Bigl(1+\frac{2t^2}{\th}\Bigr)t^2-
\frac{2}{\th}\Bigl(x^2-y^2\Bigr)t^2 - \frac{4}{\th^2}y\,r^2t^3\right \}\quad.
\end{equation}
By determining  maximum values of this function one can see the ring-like 
structure of the solution and find the dependence of their size and shape 
on $t$.  Note that these rings need not be symmetric.
We see that our field (\ref{Phi}) describes a
two-soliton configuration, with both solitons sitting at the point $x{=}y{=}0$
and changing their shape with time.

In order to construct more general time-dependent multi-solitons
with a ring structure, one may start from a seed solution 
parametrized by (higher-rank) projectors
$P_k=|0\>\<0|+\ldots+|k\>\<k|$ or 
$P_k'=\sum^k_{i,j=1}|\a_i\>\frac{1}{\<\a_i|\a_j\>}\<\a_j|$, 
where $|\a_i\>=\e^{\a_i a^\dagger-\bar\a_i a}|0\>$ denote coherent states. 
In this context, we did not find {\it abelian soliton scattering\/} 
in the modified sigma model in $2{+}1$ dimensions.
To accomplish this, one should perhaps begin with time-dependent seed 
configurations or use a more general ansatz featuring $\tilde\psi$ with 
third-order poles in~$\zeta$.

\section{Concluding remarks}

The present paper has applied a solution-generating technique 
(the dressing method) to the noncommutative modified $U(n)$ sigma model 
in $2{+}1$ dimensions.  
In this way we have constructed time-dependent classical field configurations
which describe {\it interacting\/} noncommutative solitons.
We have obtained ring-shaped bound states as well as nontrivial scattering 
solutions, albeit the latter ones only for the nonabelian case.
It would be interesting to construct the Seiberg-Witten map for
our soliton configurations as it was done e.g. in~\cite{hashi,aschieri}
for scalar solitons. Also, the brane interpretation of these solitons
has not been completely clarified.

Our construction is based on so-called dressing transformations, which map 
(old) solutions into (new) solutions. 
In fact, the set of all such transformations forms an infinite-dimensional 
group which acts on the space of all solutions of an integrable model 
(see e.g.~\cite{Uh,BB} for discussion and references). 
Multi-solitons parametrize finite-dimensional orbits of this dressing group.

As was explained in two previous papers~\cite{LPS1,LPS2}, the massless 
modes of the open \N2 string~\footnote{
For a review see~\cite{marcus,dubna}.} 
in a space-time filling brane with a constant NS $B$-field 
are described by noncommutative self-dual Yang-Mills (SDYM) theory 
in $2{+}2$ dimensions. Upon reduction on a D2-brane worldvolume, 
the modified sigma model under consideration emerges.
Therefore, the algebra of all (nonlocal) dressing
symmetries of our field equations can be obtained (after noncommutative
generalization) from the dressing symmetries of the SDYM equations (see 
e.g.~\cite{PP,Po2} and references therein) or from their stringy 
generalization~\cite{ivle,pole3}. 
Moreover, not only the modified sigma-model equations but a lot of other 
integrable equations in three and fewer dimensions derive from the SDYM
equations by suitable reductions (see e.g.~\cite{ward3,MS,IP,legare2,dimakis2}
and references therein). The investigation of the corresponding reductions of 
the {\it noncommutative\/} SDYM equations to noncommutative versions of KdV, 
nonlinear Schr\"odinger and some other integrable equations has just begun
\cite{taka,legare,dimakis,paniak}. 
Their solitonic solutions are worth exploring.

\bigskip
\noindent
{\large{\bf Acknowledgements}}

\smallskip
\noindent
We thank M. Wolf for careful reading of the manuscript and 
pointing out several typos.

\end{document}